\documentstyle[preprint,aps,prc]{revtex}

\begin{document}

\draft

\title{Superheavy nuclei in selfconsistent nuclear calculations}

\author{K. Rutz${}^{a}$, M. Bender${}^{a}$, 
        T. B\"urvenich${}^{a}$, T. Schilling${}^{a}$,
        P.--G. Reinhard${}^{b,c}$, J. A. Maruhn${}^{a,c}$,
        and W. Greiner${}^{a,c}$}

\address{${}^{a}$ Institut f\"ur Theoretische Physik, 
         Universit\"at Frankfurt\\
         Robert-Mayer-Str. 10, D-60325 Frankfurt, Germany.}
\address{${}^{b}$ Institut f\"ur Theoretische Physik, 
         Universit\"at Erlangen\\
         Staudtstr.\ 7, D-91058 Erlangen, Germany}
\address{${}^{c}$ Joint Institute for Heavy-Ion Research, 
         Oak Ridge National Laboratory\\
         P. O. Box 2008, Oak Ridge, TN 37831, U.S.A.}

\date{22 October 1996}

\maketitle

\begin{abstract}
The shell structure of superheavy nuclei is investigated within 
various parametrizations of relativistic and nonrelativistic 
nuclear mean field \mbox{models}. The heaviest known even-even nucleus 
${}^{264}_{156}{\rm Hs}_{108}$ is used as a benchmark to estimate 
the predictive value of the models. From that starting point, 
doubly magic spherical nuclei are searched in the region 
$Z\!=\!110\!-\!140$ and $N\!=\!134\!-\!298$. They are found at
($Z\!=\!114\,,\,N\!=\!184$), ($Z\!=\!120\,,\,N\!=\!172$), or at 
($Z\!=\!126\,,\,N\!=\!184$), depending on the parametrization.
\end{abstract}

\pacs{PACS numbers: 21.60.Jz, 21.30.Fe, 24.10.Jv, 27.90.+b}

\section{Introduction}
The possible existence of islands of shell-stabilized superheavy 
nuclei has been an inspiring problem in heavy-ion physics for 
almost three decades \cite{RevSH}. Recent experiments at GSI 
\cite{Z111,Z112} and Dubna \cite{Dubna} brought innovations by 
producing isotopes at 
and in the vicinity of the {\em deformed} doubly magic nucleus
${}^{270}_{162}{\rm Hs}_{108}$, as theoretically verified in 
macroscopic-microscopic models \cite{Patyk,FY}. The ultimate
goal remains 
the spherical doubly magic superheavy nucleus ${}^{298}_{184}114$
which was predicted in the earliest macroscopic-microscopic
investigations \cite{Mosel,SuperNils} and confirmed in more recent
models of this type \cite{Patyk,FY}. The expectation that in near 
future experimental progress will access this region is a strong 
motivation to investigate the shell structure of superheavy nuclei
within the self-consistent nuclear mean-field models \cite{Naz}, 
especially since there were early indications \cite{SHSIII} that 
proton and neutron shell closures strongly affect each other and 
that $Z\!=\!120$ may be a shell closure. 

It is the aim of this contribution to scan a wide region of 
superheavy nuclei for occurence of
spherical magic shells within the framework of the relativistic 
mean-field model (RMF) (for reviews see \cite{WalSer,Rei89}) 
and within the nonrelativistic Skyrme--Hartree--Fock (SHF) 
approach (for a review see \cite{refSHF}).

The extrapolation towards superheavy nuclei challenges the 
predictive power of nuclear structure models. The 
macroscopic-microscopic method, although generally successful, 
requires preconceived knowledge about the expected densities 
and single-particle potentials, which fades away when
stepping into new regions where stronger polarization effects 
and more complicated functional forms of the densities may occur. 
These effects are naturally
incorporated within selfconsistent nuclear models which nowadays 
manage to describe all known nuclei from $^{16}{\rm O}$ on with 
satisfying quality by fixing a handful of model parameters 
\cite{Rei89,SLyx,SkIx}.
There remain, however, several loosely fixed aspects in these
parametrizations which amplify as uncertainties in extrapolations, 
e.g., to nuclei near the drip line \cite{drip} or to superheavy 
nuclei as discussed here. 

\section{The framework}

In view of the uncertainties, we consider a broad selection of
parametrizations with about comparable quality concerning normal 
nuclear properties but differences in some detail.

For the nonrelativistic SHF calculations we consider the
parametrizations ${\rm SkM}^{*}$ \cite{SkM*}, SkI1 \cite{SkIx}, SkP
\cite{SkP}, SLy6 \cite{SLyx} which all employ the standard form
but differ in bias. The force SkP uses effective mass 
$m^*/m\!=\!1$ and is designed to allow a selfconsistent treatment 
of pairing. The other forces all have smaller effective masses 
around $m^*/m\!=\!0.7\!-\!0.8$. The force ${\rm SkM}^{*}$ was 
first to deliver acceptable incompressibility and fission 
properties and it is still a benchmark in this area. The 
force SLy6 stems from an attempt to cover properties 
of pure neutron matter together with normal nuclear ground state
properties; one can expect reliable extrapolations to neutron rich
nuclei from this force. The force SkI1 stems from a recent 
systematic fit (along the strategy of \cite{SkyrmeFit}) already 
embracing data from exotic nuclei; it is biased towards an optimal
description of normal nuclei including surface properties.

The forces SkI3 and SkI4 are fitted exactly as SkI1 but using 
a variant of the Skyrme parametrization where the spin-orbit force 
is complemented by an explicit isovector degree-of-freedom 
\cite{SkIx}. They are designed to overcome the different isovector
trends of spin-orbit coupling
between conventional Skyrme forces and the RMF. SkI3 contains a
fixed isovector part exactly analogous to the RMF, whereas SkI4 is
adjusted allowing free variation of the isovector spin-orbit force.
Both forces contain a minimal relativistic correction within 
the SHF ansatz. The modified spin-orbit force has a strong effect 
on the spectral distribution in heavy nuclei and we expect visible 
consequences for the predictions of superheavy nuclei.

For the RMF we consider the parametrizations NL-Z \cite{NLZ}, PL-40
\cite{PL40}, NL-SH \cite{NLSH}, and TM1 \cite{TM1}. The force NL-Z 
aims
at a best fit to nuclear ground state properties for the standard
nonlinear ansatz \cite{Rei89} with cubic and quartic selfcoupling
of the scalar field. The force PL-40 is a similar fit, but with a
stabilized form of the scalar nonlinear selfcoupling. It shares 
most properties with NL-Z, as the good reproduction of ground state
properties and similar nuclear matter properties with the low 
effective mass $m^*/m=0.58$ which is typical for the RMF. But 
PL-40 is somewhat more appropriate in the regime of small 
densities at the outer nuclear surface and thus yields better 
fission barriers \cite{Asymrmf}. The force NL-SH also employs the 
standard ansatz, but was adjusted with a bias to exotic nuclei, 
fitting neutron radii instead of surface thicknesses.
Finally, the force TM1 includes a nonlinear selfcoupling of the 
vector field as well, and is fitted in the same way as NL-SH.

In both, SHF and RMF, the pairing correlations are treated in 
the BCS scheme using a delta pairing force \cite{SLyx} 
$V_{\rm pair} = V_{\rm p/n} \, \delta(r_1 - r_2)$. The strengths
$V_{\rm p}$ for protons and $V_{\rm n}$ for neutrons depend
on the actual mean--field parametrization. They are optimized by
fitting (for each parametrization separately) the pairing gaps 
in Sn isotopes and the isotones with $N \! = \! 82$. The pairing 
space was chosen twice as large as the given particle number 
with a smooth Fermi cutoff weight, for details see 
\cite{pairStrength} 

Furthermore, a center-of-mass correction is employed, for the 
SkI$x$, SLy6, NL-Z and PL-40 forces by substracting a posteriori
$E_{\rm c.m.}=\langle\hat{P}^2_{\rm c.m.}\rangle/2mA$, for NL-SH 
and TM1 by substracting the harmonic oscillator estimate 
$E_{\rm c.m.}=\frac{3}{4} \; 41 A^{-1/3} \; {\rm MeV}$, 
while for SkM* and SkP only a diagonal correction is performed 
\cite{SkyrmeFit}, as used in the original adjustment of these 
parameter sets.

The numerical procedure solves the coupled SHF and RMF equations 
on a grid in coordinate space with the damped gradient iteration 
method \cite{dampgrad}. A spherical representation is employed 
in most of the calculations. An axially symmetric deformed 
representation has been used occasionally for counterchecks and 
particularly for the deformed system ${}^{264}{\rm Hs}$.

To summarize the features of the forces subject to our 
investigation: All provide about the same good quality concerning 
the nuclear bulk properties, energies and radii, in known stable 
nuclei. There are differences in surface properties: Most forces 
perform very well in that respect, but the forces NL-SH and TM1 
produce a too small surface thickness and correspondingly do 
not work so well in fission calculations; this holds, although 
less dramatically, for the force SLy6. There are differences in 
the effective mass: The modern fits SkI$x$,
SLy6, NL-Z, and PL-40 all have low effective masses (below 0.7 for 
SHF and below 0.65 for the RMF models) whereas SkP even comes up to
$m^*/m\!=\!1$; this has consequences on the level density and
thus on the shell structure in large systems. There are 
differences in the description of neutron rich nuclei: the
forces NL-SH and SLy6 are especially designed for this aspect, the
forces SkI$x$ include some information from the neutron rich area
in their fit, and the performance of all the other forces in 
that respect is yet untested. There are differences in isotopic 
trends of radii: all genuine SHF forces fail in that respect 
whereas RMF models do very well; the forces SkI3 and SkI4 use 
an extended Skyrme ansatz which cures that defect and also provide 
good isotopic trends. In that respect SkI4 is superior.
Table \ref{NucMat} summarizes the nuclear matter properties 
of the forces discussed and gives an overview of the reproduction 
of the isotope shifts on charge radii in lead and the spin-orbit 
splitting in ${}^{16}{\rm O}$.

\section{Comparison for an existing superheavy nucleus}

The question is now how all these parametrizations, which provide 
nearly comparable quality in the regime of known stable nuclei 
but differ in some details perform when extrapolating to the 
new area of superheavy nuclei. Before going into the regime 
of the yet unknown, we therefore take the presently heaviest known
nuclei as benchmarks. To that
end we have calculated the ground states of the heaviest even-even
nucleus for which the mass is known, i.e.
${}^{264}_{156}{\rm Hs}_{108}$ \cite{Wapstra}. This nucleus is 
close to a region of enhanced stability in the vicinity of the 
doubly deformed magic nucleus ${}^{270}_{162}{\rm Hs}_{108}$ 
\cite{Patyk,FY}.

Table~\ref{Hs264} shows ground state properties of
${}^{264}{\rm Hs}$ obtained from deformed mean field calculations 
for the variety of forces explained above. The experimental 
binding energy is also given for comparison.
The dimensionless multipole deformations are defined as
$\beta_\ell=4\pi\langle r^\ell Y_{\ell0}\rangle / (3Ar_0^\ell)$
with $r_0 = 1.2\,A^{1/3}$
and provide a more immediate geometrical understanding than the
multipole moments as such \cite{beta}. We see from 
table~\ref{Hs264}
that almost all models agree in the predicted deformations, which
corroborates the experience that well developed deformations are a
general topological feature of nuclear shell structure
\cite{SuperNils,funnyhills}. There is, however, a noteworthy
exception in that NL-SH and TM1 produce a somewhat smaller 
quadrupole moment. It seems that their smaller surface thickness and 
larger effective mass modifies the shell structure 
so much that deformation properties are shifted. This 
feature is also found in fission barriers \cite{Asymrmf} and 
a systematic variation of the effective mass in studies 
of deformation energy surfaces \cite{Blum}.

The most interesting observable for our purposes is the binding 
energy, because experimental information is available.
For better comparison, in the third column we display 
the relative errors between calculation and experimental value. 
Although all forces in our selection show acceptable quality in 
that extrapolated result, there are clearly visible differences.
The Skyrme forces with the old standard spin-orbit coupling have 
about the same error of about $0.6$, with recent fits coming a bit
closer than older forces. The isovector-extended spin-orbit 
coupling in SHF produces a big step forward in quality concerning 
this observable, which shows that there is some thruth in the 
relativistic isovector mix of the spin-orbit coupling. This is 
corroborated by the equally good results of the RMF forces NL-Z 
and PL-40. There is, however, a different sign in the error which 
hints at an essential difference between SHF and RMF, yet to be 
understood. The "exotic" RMF forces NL-SH and TM1 again
fall below the quality of the more standard parametrizations. The
conclusion from table~\ref{Hs264} is that for the extrapolations to
superheavy nuclei, the forces SkI3, SkI4, NL-Z, and PL-40 should be
preferred.

\section{Spherical magic shells in larger superheavy nuclei}

The most interesting feature for even larger systems is the 
possible occurence of new spherical doubly magic nuclei. There are
different possibilities to identify magic numbers. One often 
considers a gap in the single-particle spectra as a signal for 
a magic number, but this is not always sufficient. 
In macroscopic-microscopic models the shell correction provides 
a natural measure for magicity. The shell correction is related 
to the difference between the experimental values of the nuclear 
masses and the predictions of a liquid-drop model. A more direct 
measure of a shell closure is the observation of a sudden jump in
the two-nucleon separation energies,
$S_{\rm 2p}(N,Z) = B(N,Z) - B(N,Z-2)$ for the protons or
$S_{\rm 2n}(N,Z) = B(N,Z) - B(N-2,Z)$ for the neutrons.
Therefore the two-nucleon gaps,
\begin{eqnarray}
 \delta_{\rm 2p} (N,Z) &=& 2B(N,Z)-B(N,Z-2)-B(N,Z+2)
\nonumber \\
 \delta_{\rm 2n} (N,Z) &=& 2B(N,Z)-B(N-2,Z)-B(N+2,Z)
\label{defgaps}
\end{eqnarray}
show a pronounced peak for magic numbers \cite{w.naz}. We will consider the
two-nucleon gaps (\ref{defgaps}) as the observable with large
positive values indicating a shell closure. The scale of this 
quantity is indicated by the gaps for the doubly-magic 
$^{208}{\rm Pb}$ which are $\delta_{\rm p}=8.5\,{\rm MeV}$ and 
$\delta_{\rm n}=7.8\,{\rm MeV}$ for SkI1. It is to be noted 
that the amplitude of
shell effects decreases with increasing system size, due to the
increasing level density. This will make it more and more difficult
to find proncounced gaps for much larger systems.

The calculations in spherical symmetry are of significance for 
experiments only where they indicate double shell closures. If 
either kind of nucleons tends to deformation, the shell structure
of other kind need not be strong enough to preserve sphericity.

Figure~\ref{figgaps} shows the proton and neutron gaps from
spherical mean-field calculations with the chosen forces for 
a large variety of $Z$ and $N$. The results from force NL-Z 
are so close to those of PL-40 that we have displayed only 
one case. As expected, the largest gaps are much smaller than 
in the lead region (by about a factor of 2). In the following 
discussion we will consider the black squares (standing for the 
largest gaps) as indicators of a shell closure. The left column of
figure~\ref{figgaps} shows the proton gaps $\delta_{\rm p}$. 
The isotopes of $Z\!=\!120$ have the most pronounced proton 
gaps in all cases, except for SkI4 where $Z\!=\!114$ is the 
preferred case, respectively SkM* and SkP, where $Z\!=\!126$ 
is favoured. 

The right column of figure~\ref{figgaps} shows the neutron gaps 
$\delta_{\rm n}$. The
dominant shell closure is $N\!=\!184$ which appears for all forces
except for NL-SH and TM1. The latter force shows no pronounced
neutron shells at all. The forces PL-40 and NL-Z, on the other 
hand, deliver even an alternative (and preferred) choice with 
$N\!=\!172$, which appears to be the dominant shell closure for 
NL-SH and TM1. Generally, it is to be noted that those four forces 
which are preferred from comparison with ${}^{264}{\rm Hs}$ 
produce the best developed shell closures for protons, whereas in
all standard SHF models (SLy6, SkI1, SkM$^*$, SkP) and the 
relativistic NL-SH as well as TM1 the shell structure appears 
to be less pronounced. The more reliable forces thus prefer shell 
closures and this hints that some magic system will be observed 
in that range of nuclei.

The most interesting species are, of course, the doubly magic 
systems. These require a  simultanous occurence of a large shell 
gap (black squares) for the protons (left column) as well as for 
the neutrons (right column). It is interesting to note that such 
a coincidence is not trivial, as we see from the many cases where 
it cannot be found (SkI1, SkI3, SLy6, TM1). The remaining 
parametrizations do predict doubly magic nuclei, however, at 
different places. The forces SkP and ${\rm SkM}^{*}$ predict 
$Z\!=\!126$, $N\!=\! 184$. The presection with ${}^{264}{\rm Hs}$ 
has picked the two forces SkI4 and PL--40 (=NL--Z) both of which 
show doubly magic nuclei. The relativistic
PL--40 parametrization predicts $Z\!=\!120$, $N\!=\!172$, whereas
the nonrelativistic SkI4 prefers $Z\!=\!114$, $N\!=\!184$. 
Thus even two optimized and preselected forces make conflicting 
predictions. It is to be noted that shell models usually predict 
the doubly magic $Z\!=\!114$, $N\!=\!184$ 
\cite{Patyk,FY,Mosel,SuperNils}. The more robust occurence of
the magic $N\!=\!184$ neutron shell and the more favourable charge
asymmetry seem to indicate a preference for this configuration.

We prefer, however, to read the result the other way round. The 
study of superheavy nuclei has disclosed significant deviations 
amongst a set of otherwise comparable mean-field models. In 
particular, there is a systematic difference between the RMF 
and SHF models which has yet to be understood. New experimental 
information on superheavy nuclei will help to clarify these open 
theoretical questions.

One sees in figure~\ref{figgaps} that the proton shell closures for
a given $Z$ can change with varying neutron number, and similarly 
the neutron shell closures vary with changing proton numbers. A 
vivid example is the $Z\!=\!120$ shell computed with SkI1 which 
starts with closure, looses that property with increasing neutron 
number, and regains it later. The changes are related to a 
changing level density at the Fermi surface. As a demonstration, 
we show in figure~\ref{figspectr} the single proton spectra for 
this case, i.e. $Z\!=\!120$ computed with SkI1. One has to
watch the shell gap at $Z\!=\!120$. Minimal relative changes of the
single proton levels indeed produce a regime of higher level 
densities around $N\!=\!184$, the neutron number where the proton 
shell gap is lowest, see figure~\ref{figgaps}. This example 
illustrates that shell closures in superheavy nuclei are an 
extremely sensitive property. It is no surprise that this 
question imposes severe constraints on models and forces.

\section{Conclusions}

We have investigated the description of superheavy nuclei in the
framework of relativistic and nonrelativistic nuclear mean-field 
models. A representative selection of parametrizations is 
considered which provide all about the same good quality 
concerning nuclear bulk observables but differ with respect to 
surface tension, effective mass, and isovector
features. We take advantage of the heaviest experimentally measured
even-even nucleus and use its binding energy to check the 
predictive power of the preselected forces. This shows a clear 
preference for the standard
relativistic forces (NL-Z, PL-40) and relativistically corrected
Skyrme forces (SkI4, SkI3). Shell closures are quantified in 
terms of the shell gap, i.e. the second difference of binding 
energies. A systematic survey of shell gaps in the range of 
$110 \! < \! Z \! < \! 140$ and
$134 \! < \! N \! < \! 298$ shows that the preferred forces also 
provide more pronounced shell closures. There remain, however, 
conflicting predictions for a doubly magic system: $Z\!=\!120$, 
$N\!=\!172$ for the relativistic forces PL-40, NL-Z and NL-SH but 
$Z\!=\!114$, $N\!=\!184$ for the nonrelativsitic force SkI4 and 
$Z\!=\!126$, $N\!=\!184$ for the standard Skyrme forces SkM* and 
SkP. Additional criteria (general trends, shell
model predictions, charge asymmetry) set a preference on the case
$Z\!=\!114$, $N\!=\!184$. But the conclusion is rather that the 
study of superheavy systems remains a challenge for selfconsistent
nuclear mean-field models, which have to be developped to a new 
stage by much more rigorous testing of a wide variety of nuclear 
properties throughtout the periodic table. 
In particular the results have revealed a systematic
difference between the relativistic and the nonrelativistic models
which deserves further close inspection.

\section*{Acknowledgements}
The authors would like to thank S.~Hofmann, G.~M\"unzenberg, and 
D.~Habs for many valuable discussions.

This work was supported by 
Bundesministerium f\"ur Bildung und Forschung (BMBF),
by Deutsche Forschungsgemeinschaft (DFG),
by Gesellschaft f\"ur Schwerionenforschung (GSI),
and by Graduiertenkolleg Schwerionenphysik.

The Joint Institute for Heavy Ion Research has as member 
institutions the University of Tennessee, Vanderbilt University, 
and the Oak Ridge
National Laboratory; it is supported by the members and by the
Department of Energy through Contract No.\ DE-FG05-87ER40361 with 
the University of Tennessee.

\newpage

\begin{table}[t]   
\caption{\label{NucMat}} 
{Compilation of nuclear matter properties
for the parameter sets used in this study. 
$E/A$ and $\rho_{0}$ denote the equilibrium energy per nucleon 
and density, $K_{\infty}$ the compression modulus, $m^{*}$ the 
effective mass (caution:
defined differently for relativistic and nonrelativistic models
\cite{Effmass}) 
and $a_{\rm sym}$ the asymmetry coefficient.
$\Delta r^2_{\rm c}$ is the isotope shift on charge r.m.s. radii 
for ${}^{214}{\rm Pb}-{}^{208}{\rm Pb}$, 
$\epsilon_{ls}$ the spin--orbit splitting between the 
$1 {\rm p}_{3/2}$ and $1 {\rm p}_{1/2}$ level in ${}^{16}{\rm O}$, 
see \cite{SkIx}.}

\begin{tabular}{lcccccccccc}
Force & $E/A \, [{\rm MeV}]$
      & $\rho_0 \, [{\rm fm}^{-3}]$
      & $K_\infty \,[{\rm MeV}]$
      & $m^*/m$
      & $a_{\rm sym}$
      & $\Delta r^2_{\rm c} \, [{\rm fm}^2]$
      & $\epsilon_{ls,{\rm p}}\, [{\rm MeV}]$
      & $\epsilon_{ls,{\rm n}}\, [{\rm MeV}]$ \\ \hline
SkM*  & $-16.01$ & 0.160 & 217 & 0.789 & 30.0 & 0.359 
& 6.2 & 6.3 \\
SkP   & $-16.04$ & 0.163 & 202 & 1.000 & 30.0 & 0.371 
& 4.5 & 4.6 \\
SLy6  & $-15.92$ & 0.159 & 230 & 0.690 & 32.0 & 0.428 
& 5.7 & 5.8 \\
SkI1  & $-15.93$ & 0.160 & 243 & 0.693 & 37.5 & 0.380 
& 6.1 & 6.2 \\
SkI3  & $-15.96$ & 0.158 & 258 & 0.577 & 34.8 & 0.567 
& 6.3 & 6.3 \\
SkI4  & $-15.92$ & 0.160 & 248 & 0.650 & 29.5 & 0.600 
& 6.3 & 6.2 \\ \hline
NL-Z  & $-16.19$ & 0.151 & 174 & 0.58  & 41.8 & 0.650 
& 5.8  & 5.8 \\
PL-40 & $-16.17$ & 0.153 & 166 & 0.58  & 41.7 & 0.698 
& 5.8  & 5.9 \\
NL-SH & $-16.33$ & 0.146 & 355 & 0.66  & 36.1 & 0.587 
& 6.8  & 6.9 \\
TM1   & $-16.3 $ & 0.145 & 281 & 0.634 & 36.9 & 0.646 
& 5.6  & 5.7 \\ \hline
exp.  &  --    &  --     &  -- & --    & --   & 0.613 
& 6.3  & 6.1 
\end{tabular}  
\end{table}  
\newpage
\begin{table}[t] 
\caption{\label{Hs264}}
{Binding energy (in units of ${\rm MeV}$), relative error on 
binding 
energy, quadrupole deformation $Q_2$ in units of ${\rm fm}^2$ and 
dimensionless quadrupole ($\beta_2$) and hexadecapole ($\beta_4$) 
deformations 
for ${}^{264}{\rm Hs}$ computed for several mean field 
parametrizations as indicated in the first column. YPE+WS is 
the result of a macroscopic-microscopic calculation \cite{Patyk}.
The last line shows the experimental binding energy from 
\cite{Wapstra}.} 

\begin{tabular}{lccccc}
Force & $E\,[{\rm MeV}]$
      & $\delta E/E \, [\% ]$
      & $Q_2 \, [{\rm fm}^2]$
      & $\beta_2$
      & $\beta_4$ \\ \hline
SkM*   & $-1907.18$ & 1.01 & 1033 & 0.28 & $-0.01$ \\ 
SkP    & $-1914.81$ & 0.61 & 1053 & 0.28 & $-0.01$ \\
SLy6   & $-1915.89$ & 0.56 & 1034 & 0.28 & $-0.02$ \\
SkI1   & $-1915.24$ & 0.59 & 1057 & 0.28 & $-0.02$ \\
SkI3   & $-1920.02$ & 0.34 & 1020 & 0.27 & $-0.02$ \\
SkI4   & $-1923.51$ & 0.17 & 1012 & 0.27 & $-0.02$ \\ \hline
NL-Z   & $-1931.32$ & $-0.24$& 1074 & 0.29 & $+0.00$ \\
PL-40  & $-1931.34$ & $-0.24$& 1072 & 0.29 & $+0.00$ \\ 
NL-SH  & $-1939.14$ & $-0.64$&  904 & 0.24 & $+0.00$ \\
TM1    & $-1938.66$ & $-0.62$&  945 & 0.25 &  0.02 \\ \hline
YPE+WS & $-1925.89$  & 0.04    & --      &  0.24  & $-0.03$ \\ \hline
Exp.   & $-1926.72$  &       &           &        & 
\end{tabular}
\end{table}

\begin{figure}
\vspace*{1cm}
\caption{\label{figgaps}}
{  Grey scale plots of proton gaps (left column) and neutron gaps
   (right column) in the $N$-$Z$ plane for spherical calculations 
   with the forces as indicated. The assignment of scales differs 
   for protons and neutrons, see the uppermost boxes where the 
   scales are indicated in units of MeV. Nuclei that are stable 
   with respect to $\beta$ decay and the two-proton dripline are 
   emphasized.}
\end{figure}

\begin{figure}
\vspace*{1cm}
\caption{\label{figspectr}}
{  The single proton levels near the Fermi energy for the isotopes
   of $Z\!=\!120$ versus the neutron number, computed with SkI1.
   Owing to minimal relative changes of the single proton levels 
   the proton gap at $Z\!=\!120$ vanishes in the vicinity of
   $N\!=\!184$, the neutron number where the proton shell gap 
   $\delta_{\rm p}$ is lowest, see figure~\ref{figgaps}. 
}
\end{figure}


\begin{thebibliography}{20}

\bibitem{RevSH}
   K. Kumar, 
   {\em Superheavy Elements}, Adam Hilger, Bristol and New York, 
    1989.

\bibitem{Z111}
   S. Hofmann, V. Ninov, F. P. Hessberger,  P. Armbruster, 
   H. Folger, 
   G. M\"unzenberg, H. J. Sch\"ott,  A. G. Popeko, A. V. Yeremin, 
   A. N. Andreyev, S. Saro, R. Janik, and  M. Leino, 
   Z. Phys. {\bf A350}, 277 (1995) and 
   Z. Phys. {\bf A350}, 281 (1995).

\bibitem{Z112}
   S. Hofmann, V. Ninov, F.P. Hessberger,  P. Armbruster, 
   H. Folger, G. M\"unzenberg, H. J. Sch\"ott,  A. G. Popeko, 
   A. V. Yeremin, S. Saro, R. Janik, and  M. Leino, 
   Z. Phys. {\bf A354}, 229 (1996).

\bibitem{Dubna}
   Yu. A. Lazarev, Yu. V. Lobanov, Yu. Ts. Oganessian, 
   V. K. Utyonkov,
   F. Sh. Abdullin, A. N. Polyakov, J. Rigol, I. V. Shirokovsky,
   Yu. S. Tsyganov, S. Iliev, V. G. Subbotin, A. M. Sukhov,
   G. V. Buklanov, B. N. Gikal, V. B. Kutner, A. N. Mezentsev,
   K. Subotic, J. F. Wild, and R. W. Lougheed, K. J. Moody,
   Phys. Rev. {\bf C54}, 620 (1996).

\bibitem{Patyk} 
   Z. Patyk and A. Sobiczewski, 
   Nucl. Phys. {\bf A533}, 132 (1991).

\bibitem{FY}
   P. M\"oller, J.~R.~Nix, 
   Nucl. Phys. {\bf A549}, 84, (1992);
   J. Phys. {\bf G 20}, 1994 1681.

\bibitem{Mosel} 
   U.~Mosel, W.~Greiner, 
   Z.~Phys.~{\bf 222}, 261 (1969).

\bibitem{SuperNils}
   S.~G.~Nilsson, C.~F. Tsang, A.~Sobiczewski, Z.~Szymanski, 
   S.~Wycech, C.~Gustafson, I.--L.~Lamm, P.~M\"oller, B.~Nilsson,
   Nucl.~Phys.~{\bf A131}, 1 (1969).

\bibitem{Naz}
   S.~\'Cwiok, J.~Dobaczewski, P.-H.~Heenen, P.~Magierski, 
   W.~Nazarewicz,
   preprint, 1996.

\bibitem{SHSIII} 
   M.~Beiner, H.~Flocard, M.~V{\'e}n{\'e}roni, P.~Quentin, 
   Physica Scripta {\bf 10A}, 84 (1974).

\bibitem{WalSer}
   B.~D.~Serot, J.~D.~Walecka, 
   Adv. Nucl. Phys. {\bf 16}, 1 (1986).

\bibitem{Rei89}
   P.--G.~Reinhard, 
   Rep.~Prog.~Phys.~{\bf 52}, 439 (1989).

\bibitem{refSHF}
   P.~Quentin, H.~Flocard, 
   Ann. Rev. Nucl. Part. Sci. {\bf 28}, 523 (1978).

\bibitem{SLyx}
   E.~Chabanat, P.~Bonche, P.~Haensel, J.~Meyer, R.~Schaeffer,
   preprint, 1996.

\bibitem{SkIx}
   P.--G.~Reinhard, H.~Flocard,  
   Nucl.~Phys.~{\bf A584}, 467 (1995).

\bibitem{drip}
   W.~Nazarewicz, J.~Dobaczewski, T.~R.~Werner, J.~A.~Maruhn, 
   P.--G.~Reinhard, K.~Rutz, 
   C.~R.~Chinn, A.~S.~Umar, M.~R.~Strayer,
   Phys. Rev. {\bf C53}, 740 (1996).

\bibitem{SkM*}
   J.~Bartel, P.~Quentin, M.~Brack, C.~Guet, H.--B.~H{\aa}kansson,
   Nucl.~Phys.~{\bf A386}, 79 (1982).

\bibitem{SkP}
   J.~Dobaczewski, H.~Flocard, J.~Treiner, 
   Nucl. Phys. {\bf A422}, 103 (1984).

\bibitem{SkyrmeFit}
   J.~Friedrich, P.--G.~Reinhard, 
   Phys. Rev. {\bf C33}, 335 (1986).

\bibitem{NLZ}
   M.~Rufa, P.--G.~Reinhard, J.~A.~Maruhn, W.~Greiner, 
   M.~R.~Strayer,
   Phys. Rev. {\bf C38}, 390 (1989).

\bibitem{PL40}
   P.--G.~Reinhard, 
   Z.~Phys. {\bf A 329}, 257 (1988).

\bibitem{NLSH}
   M.~M.~Sharma, P.~Ring, 
   Phys. Rev. {\bf C45}, 2514 (1992).

\bibitem{TM1}
   Y.~Sugahara, H.~Toki, 
   Nucl. Phys. {\bf A579}, 557 (1994).

\bibitem{Asymrmf}
   K.~Rutz, J.~A.~Maruhn, P.--G.~Reinhard, W.~Greiner,
   Nucl.~Phys.~{\bf A590}, 680 (1995).

\bibitem{pairStrength}
   M.~Bender, P.--G.~Reinhard, K.~Rutz, J.~A.~Maruhn,
   (to be submitted). 

\bibitem{dampgrad}
   V.~Blum, G.~Lauritsch, J.~A.~Maruhn, P.--G.~Reinhard,
   J. Comp. Phys. {\bf 100}, 364 (1992).

\bibitem{Wapstra} 
   G.~Audi and A.~H.~Wapstra, 
   Nucl. Phys. {\bf A595}, 409 (1995).  

\bibitem{beta} R.~W.~Hasse and W.~D.~Myers,  
   {\sl Geometrical Relationships of Macroscopic Nuclear Physics},
   Springer, Berlin, 1988.

\bibitem{funnyhills}
   M.~Brack, J.~Damg{\aa}rd, A.~S.~Jensen, H.~C.~Pauli, 
   V.~M.~Strutinsky, C.~Y.~Wong, 
   Rev.~Mod.~Phys.~{\bf 44}, 320 (1972).
                                               
\bibitem{Blum}
   V.~Blum, J.~A.~Maruhn, P.--G.~Reinhard, W.~Greiner,
   Phys.~Lett. {\bf B323}, 262 (1994).

\bibitem{w.naz}
   W.~Nazarewicz, J.~Dobaczewski, T.~R.~Werner,
   Physica Scripta {\bf T56}, 9 (1995).

\bibitem{Effmass} 
   M.~Jaminon, C.~Mahaux, 
   Phys. Rev. {\bf C40}, 354 (1989).

\end{thebibliography}
\end{document}